\documentclass[pre,twocolumn,showpacs,floatfix]{revtex4}
\usepackage{graphicx}

\def\be{\begin{equation}}
\def\ee{\end{equation}}

\begin{document}
\title{The conditional mean acceleration of fluid particle
in developed turbulence}
 \author{A.K. Aringazin}
 \email{aringazin@mail.kz}
 \altaffiliation[Also at ]
 {Department of Mechanics and Mathematics, Kazakhstan Division,
 Moscow State University, Moscow 119899, Russia.}
 \affiliation{Department of Theoretical Physics, Institute for
Basic Research, Eurasian National University, Astana 473021
Kazakhstan}

\date{27 January 2004}

\begin{abstract}
Using the random intensity of noise (RIN) approach to the
one-dimensional Laval-Dubrulle-Nazarenko type model for the
Lagrangian acceleration in developed
turbulence
[cond-mat/0305186, cond-mat/0305459] we study the probability
density function and mean acceleration conditional on velocity
fluctuations. The additive noise intensity and the cross
correlation between the additive and multiplicative noises are
assumed to be dependent on velocity fluctuations in an exponential
way. The obtained fit results are found to be in a good
qualitative agreement with the recent experimental data on the
conditional acceleration statistics by Mordant, Crawford, and
Bodenschatz. The fit to the observed conditional mean acceleration
is of pure illustrative character which is performed to study
influence of variation of the cross correlation parameter on the
shape of conditional acceleration distribution and conditional
acceleration variance. The conditional mean acceleration should be
zero for homogeneous isotropic turbulence. The observed
conditional mean acceleration increases for bigger velocity
fluctuation amplitude and is associated to anisotropy of the
studied flow.
\end{abstract}


\maketitle

\section{Introduction}\label{Sec:Introduction}

The nonextensive statistics developed by Tsallis~\cite{Tsallis}
has motivated a phenomenological
approach~\cite{Johal,Aringazin,Beck3} which was recently
used~\cite{Beck,Beck4} to describe Lagrangian acceleration of
fluid particle in developed three-dimensional turbulent flow
within the framework of Langevin type equation; see
also~\cite{Sawford,Wilk,Beck2,Reynolds}. Some recent stochastic
particle models and refinements of this
technique~\cite{Aringazin2,Aringazin3,Aringazin4} were reviewed in
Ref.~\cite{Aringazin5}. The one-dimensional Langevin toy models of
Lagrangian turbulence lacks of physical interpretation, {\it
e.g.,} of short term dynamics, or small-scale and large-scale
contributions, in the context of three-dimensional incompressible
Navier-Stokes equation, which is generally assumed to be capable
to describe fully developed turbulence. A deep physical analysis
of this problem has been made by Gotoh and
Kraichnan~\cite{Kraichnan0305040}.

Recently~\cite{Aringazin5,Aringazin6} we have shown that the
one-dimensional Laval-Dubrulle-Nazarenko (LDN) type toy
model~\cite{Laval,Laval2} of the acceleration dynamics with the
model turbulent viscosity $\nu_{\mathrm t}$ and coupled
delta-correlated Gaussian multiplicative and additive noises is in
a good agreement with the recent high-precision experimental data
on acceleration
statistics~\cite{Bodenschatz,Bodenschatz2,Mordant0303003}.
Particularly, we have demonstrated that the predicted contribution
to fourth order moment of acceleration {does} peak at the same
values as the experimental curve, in contrast to predictions of
most of the other stochastic particle
models~\cite{Beck,Beck4,Aringazin3,Reynolds}. We note however that
characteristic parts of the acceleration distribution are the core
and tails, while the intermediate range of accelerations which is
responsible for positions of the peaks may be not robust.

Below we discuss at some length on foundations of the present
model, and focus on modeling acceleration statistics conditional
on velocity fluctuations.

The original one-dimensional LDN model was formulated both in the
Lagrangian and Eulerian frameworks for small-scale velocity
increments. It is based on a stochastic kind of Batchelor-Proudman
rapid distortion theory (RDT) approach to the three-dimensional
Navier-Stokes equation and Gabor transformation~\cite{Laval}.
Small scales are separated and assumed to be stochastically
distorted by much larger scales. Such nonlocal interactions which
can be viewed as elongated triads are featured by the
three-dimensional LDN model, to which the one-dimensional model
makes an approximation, particularly via mimicking relationship
between stretching and vorticity. Local interactions are modeled
by the turbulent viscosity which is taken due to the
renormalization group approach.

In the present paper, we use the Lagrangian formulation of the
one-dimensional LDN model, which is characterized by a simple
structure, as an ansatz to formulate Langevin type equation for
the component of fluid particle acceleration in statistically
homogeneous and isotropic developed turbulence.

In general the Langevin type equation which we employ here
contains wellknown terms, namely, time derivative of the variable,
an additive noise, and a nonlinear drift term with a
multiplicative noise; see Eq.~(\ref{LangevinLaval}) below. The
noises represent large scales and are treated independent on small
scales. In the Lagrangian frame, they are taken in the simplest
realization, Gaussian white-in-time with zero mean, and are
assumed to be delta-correlated to each other. Such choice of the
noises correspond to the zero-time approximation of small
time-scale correlations viewed along the Lagrangian trajectory of
fluid particle. Direct account for finite time correlations in a
Langevin type equation is important but this is beyond the scope
of the present study, in which we concentrate on the longtime
behavior and qualitative analysis. Longtime dynamics of large
scales is ignored in this particular case of the LDN model.


In the conventional approach, wandering of the tracer particle can
be described within the framework of generalized Brownian like
motion, with the acceleration chosen to be the random variable
depending on time $t$ (a stochastic process). The statistical
isotropy allows one to decompose the acceleration into the
`transverse' and `longitudinal' components with respect to the
corresponding velocity that reduces formally the analysis to
effective two-dimensional consideration of the essentially
three-dimensional system. The isotropy is assumed to be local, in
the neighbor of fluid particle. In one-dimensional models one
considers one of the two components not coupled to each other.
Joint dynamics of the components of acceleration is of much
interest and can be considered elsewhere.


It should be emphasized here that the usual Eulerian framework
(fixed probe) represents the structural viewpoint which is
extensively studied in the literature while the Lagrangian
framework is associated to stochastic dynamical consideration, a
statistical viewpoint. These two basic approaches are different
both in their theoretical formulations and experimental technique,
and compliment each other. In the Eulerian frame, the
acceleration,
$a_i={dv_i}/{dt}\equiv\partial_tv_i+v_k\partial_kv_i$, is
expressed in terms of the velocity field and its temporal and
spatial derivatives, while in the Lagrangian frame the
acceleration is $a_i=\partial_t v_i$, in terms of the Lagrangian
velocity, $v_i=\partial_t x_i$, of fluid particle with the
coordinates $x_i(x_k(0),t)$; $i,k=1,2,3$. In the present paper we
will use the Lagrangian framework.

Measurement of the time series $x_i(t)$ for individual tracer
particle in a turbulent flow to a high resolution requires very
high speed imaging sensors, and allows one to compute its
instantaneous Lagrangian velocity and Lagrangian
acceleration~\cite{Bodenschatz}. Experimental data are collected
for such individual particles moving in the approximately
isotropic and homogeneous turbulent flow domain.

With delta-correlated noises the Langevin type model of one of the
two components of acceleration is Markovian (no memory effects) so
that the well established Fokker-Planck approximation can be used
to link the dynamical framework to a statistical (PDF) approach.
Thus, one can study stationary solution of the associated
Fokker-Planck equation for the acceleration one-point probability
density function $P(a,t)$ under the assumption of balance between
the energy injected by driving forces at large scales and the
energy dissipated by viscous processes at small scales (a
statistically steady state).

In the case when stationary probability distribution can be found
exactly one can make further analysis without referring to
dynamics. The resulting stationary probability density function of
the component of acceleration, $P(a)$, contains some free
parameters inherited from the Langevin type equation. This PDF is
the main prediction of the model which can be directly fitted to
experimental data or direct numerical simulation (DNS) of the
Navier-Stokes equation.

The Random Intensity of Noise (RIN)
approach~\cite{Aringazin5,Aringazin6} makes a simple extension of
the above Langevin type model of the Lagrangian acceleration. The
main idea of this approach is to use the presence of two well
separated characteristic time scales (Kolmogorov time scale and
the integral time scale) of the system and assume that parameters
entering the resulting acceleration PDF, such as intensities of
the noises, are not constant but fluctuate at {\em large} time
scale and depend on Lagrangian velocity fluctuations.

Effectively, we approximate the time evolution of acceleration by
separating fast and slow time varying parts, and thus account for
large time scales which are usually ignored. This improves the
delta-correlation approximation of the noises adopted in the
one-dimensional LDN type model mentioned above, provided that
characteristic longtime fluctuations along the particle trajectory
occur mainly at time scales much bigger than the dissipative time
scale, typically at the integral time scale.

In general, one can assume hierarchy of a number of characteristic
time scales associated to the discrete cascade picture with
characteristic times of the flow modes of descending eddies.
However, in the present paper we simplify the consideration by
extending the large time scale up to the Lagrangian integral time
scale in order to make it more analytically tractable, in accord
to the presence of two basic characteristic scales in the
Kolmogorov 1941 (K41) picture of fully developed turbulence, and
the recent Lagrangian experimental data~\cite{Pinton}.

In the context of stochastic equation the delta-correlated
multiplicative noise provides short-time acceleration bursts the
origin of which is thought to be due to the presence of very
intense vortical structures relatively slowly varying in time.
These structures make an essential contribution to the
acceleration statistics, namely, to the tails of acceleration
probability density function, making it highly non-Gaussian.
Accounting for longtime fluctuations of the parameters corresponds
to an accounting for fluctuations of intensity of the vortical
structures, intensity of noisy incoherent background, and their
cross correlation. The key point is then to identify distributions
of the parameters. We will discuss it below and further in
Sec.~\ref{Sec:ConditionalMean}.

It is worthwhile to mention that accounting for the fluctuating
parameter $\beta$ in the chi-square Langevin model characterized
by the sole delta-correlated Gaussian additive noise~\cite{Beck}
was recently found~\cite{Aringazin5} to yield a stationary
probability density function of the same power law form as in
simple Langevin model characterized by the delta-correlated
Gaussian additive {\em and multiplicative} noises with constant
parameters and linear drift term. In other words, the chi-square
distributed fluctuations of $\beta$ appears to be mimicking the
presence of the multiplicative noise in this particular case.

In the first approximation the velocity statistics in the
Lagrangian frame is taken as usually stationary and Gaussian,
partially because one can easily deal with it analytically. Due to
the recent Lagrangian experiments~\cite{Bodenschatz} the
distribution of Lagrangian velocity is approximately Gaussian for
both the $x$ and $z$ components (the flatness is 3.2 and 2.8
respectively as compared with 3 for a Gaussian) and their
characteristic time variation is of the order of Kolmogorov time,
yet velocity fluctuations exceed root-mean square (rms) velocity
that is usually associated to large scales. This can be understood
as a manifestation of the importance of nonlocal (inter-scale)
interactions when considering small scales. It is important to
note here that the Lagrangian velocity autocorrelation, as well as
the autocorrelation of {\em absolute} value of velocity increments
in time, was found to cross zero at the integral time scale, while
the {\em full signed} velocity increments in time decay at the
Kolmogorov time scale~\cite{Pinton}.

Under the assumption that the parameters are given independent
random variables (large scales are weakly affected by small scales
in a three-dimensional high-Reynolds-number flow since the former
are local in the wave number space) characterized by stationary
statistics, the stationary probability density function obtained
from the Fokker-Planck equation makes a sense and is treated as
the distribution {\em conditional} on small scale velocity
fluctuations through assumed dependencies of the parameters on
$u$.

By this way the RIN approach enables one to study acceleration
statistics {conditional} on velocity fluctuations,
$P(a|u)=P(a|\mathrm{Parameters}(u))$. The velocity fluctuations
(large scales) are assumed to be decoupled from the acceleration
(small scales) at high Taylor microscale Reynolds numbers,
$R_\lambda>500$~\cite{Bodenschatz}.

It should be noted that such an approach is in agreement with the
Heisenberg-Yaglom picture of developed turbulence which relates
statistics of the fluid particle acceleration to statistics of
velocity fluctuations on the basis of K41 scaling theory, with
pressure gradient contribution to the acceleration variance
strongly dominating over that of viscous forces, in the inertial
range. The long-standing Heisenberg-Yaglom scaling, $\langle
a^2\rangle \sim {\bar u}^{9/2}$, where $\bar u$ is rms velocity,
was recently confirmed to a high accuracy for seven orders of
magnitude in $\langle a^2\rangle$, and was found to be broken for
$R_\lambda <500$~\cite{Bodenschatz} due to increasing coupling of
the acceleration to large scales of the flow.

Also, it should be mentioned that a similar approach, with the
variance of intermittent variable viewed phenomenologically as a
parameter which follows log-normal distribution, was considered by
Castaing, Gagne, and Hopfinger~\cite{Castaing}. From this point of
view RIN models can be referred to as Castaing type models.
However, the Castaing model does not refer to a stochastic
dynamical approach, which is an important ingredient of the
Lagrangian modeling, and does not relate the fluctuating parameter
to velocity fluctuations.

In the RIN approach the marginal (i.e., unconditional) stationary
probability density function $P(a)$ is found simply by integrating
out $u$ in the conditional distribution $P(a|u)$ with (Gaussian)
distributed $u$. This procedure requires prior determination of
dependencies of the parameters on velocity fluctuations $u$.

Particularly, in the RIN approach to the LDN type model the
assumption that the additive noise intensity $\alpha$ depends on
absolute value of velocity fluctuations $u$ in an {\em
exponential} way, $\alpha \simeq e^{|u|}$, was
found~\cite{Aringazin5} to imply a set of conditional probability
density functions $P(a|u)$ and the conditional acceleration
variance $\langle a^2|u\rangle$ which are in a good qualitative
agreement with the recent experimental data on the conditional
statistics of the $x$ component of acceleration obtained by
Mordant, Crawford, and Bodenschatz~\cite{Mordant0303003}, for the
normalized velocity fluctuations $|u|/\langle u^2\rangle^{1/2}$
ranging from 0 to 3. This issue will be discussed further in
Sec.~\ref{Sec:ConditionalMean}.

The effect of nonzero cross correlation $\lambda$ between the
additive and multiplicative noises (which models the relationship
between small-scale stretching and vorticity~\cite{Laval} and can
be seen as a small skewness of the probability density function of
the `longitudinal', i.e., pointed along the Lagrangian velocity,
component of acceleration) has been studied in
Ref.~\cite{Aringazin6} for an illustrative purpose. The
experimental unconditional acceleration probability density
function reveals small skewness that has been fitted by using the
value $\lambda=-0.005$.

The observed very small skewness of the probability density
functions for the $x$ component of acceleration can be assigned to
anisotropy of the studied $R_\lambda=690$
flow~\cite{Mordant0303003,Aringazin6} rather than to the effect of
correlation between stretching and vorticity. Particularly, it was
found that, in addition to the dependence $\alpha=\alpha(u)$, the
parameter $\lambda$ should also depend on velocity fluctuations to
meet the experimentally observed appreciable increase of the
conditional mean acceleration, $\langle a|u\rangle$, with the
increase of $|u|$~\cite{Mordant0303003}. The form of functional
dependence $\lambda(u)$ was assumed to produce a weak effect but
it was not specified.

In the present paper we fill this gap and demonstrate that (rather
strong) exponential dependence, $\lambda \simeq e^{|u|}$, is in a
good qualitative agreement with the experimental data. This fit is
of a pure illustrative character since the observed nonzero
conditional mean acceleration is due to the anisotropy of the
studied flow. We perform this fit to study influence of variation
of the cross correlation parameter on the shape of conditional
acceleration distribution and conditional acceleration variance.

The layout of the paper is as follows. In Sec.~\ref{Sec:LDN} we
outline results of the one-dimensional LDN type model of the
component of acceleration. In Sec.~\ref{Sec:ConditionalMean} we
study the conditional mean acceleration using the RIN extension of
the LDN type model, with the additive noise intensity $\alpha
\simeq e^{|u|}$, the cross correlation parameter $\lambda \simeq
e^{|u|}$, and the other parameters of the model fixed. We
summarize the obtained results in Sec.~\ref{Sec:Conclusions}.

\section{The LDN type Langevin model}
\label{Sec:LDN}

In this Section, we give only a brief sketch of the LDN model and
refer the reader to Refs.~\cite{Laval,Aringazin5} for more
details; see also recent paper~\cite{Dubrulle0304035} for the
stochastic RDT approach.

We use the exact result for probability density function of the
LDN type model obtained as a stationary solution of the
Fokker-Planck equation associated to the one-dimensional Langevin
equation for the component of
acceleration~\cite{Laval,Aringazin5},
\be\label{LangevinLaval}
\partial_t a = (\xi - \nu_{\mathrm t}k^2)a + \sigma_\perp.
\ee
This equation is a Lagrangian description in the scale space, in
the reference frame comoving with the wave number packet. This toy
model can also be viewed as a passive scalar in a compressible
one-dimensional flow~\cite{Laval}. Here, $\xi$ and $\sigma_\perp$
model stochastic forces in the Lagrangian frame and are chosen to
be Gaussian white-in-time noises,
\begin{eqnarray}\label{noises}
\langle\xi(t)\rangle=0, \
\langle\xi(t)\xi(t')\rangle = 2D\delta(t-t'), \nonumber \\
\langle\sigma_\perp(t)\rangle = 0, \
\langle\sigma_\perp(t)\sigma_\perp(t')\rangle = 2\alpha\delta(t-t'), \\
\langle\xi(t)\sigma_\perp(t')\rangle = 2\lambda\delta(t-t'),
\nonumber
\end{eqnarray}
where the averaging is over ensemble realizations. The probability
density function was calculated exactly~\cite{Aringazin5} and
appeared to be of rather complicated form despite the simplicity
of the starting equation (\ref{LangevinLaval})-(\ref{noises}). It
is given by
\be\label{PLaval} P(a) = \frac{C \exp[-{\nu_{\mathrm
t}k^2}/{D}+F(c)+F(-c)]} {(Da^2\!-\!2\lambda a
\!+\!\alpha)^{1/2}(2Bka+\nu_{\mathrm t}k^2)^{{2B\lambda
k}/{D^2}}},
\ee
where we have denoted
\begin{eqnarray}\label{A4}
F(c)
 = \frac{c_1k^2}{2c_2D^2c}\ln[\frac{2D^3}{c_1c_2(c-Da+\lambda)}
   \nonumber \\
   \times(
   B^2(\lambda^2 + c\lambda-D\alpha)a
   + c(D\nu_{\mathrm t}^2k^2+c_2\nu_{\mathrm t})
    )
   ],\\
c=-i\sqrt{D\alpha-\lambda^2}, \quad \nu_{\mathrm t} =
\sqrt{\nu_0^2+ B^2a^2/k^2},\\
c_1 = B^2(4\lambda^3\!+\!4c\lambda^2\! -\!
3D\alpha\lambda-cD\alpha)
    \!+\! D^2(c\!+\!\lambda)\nu_0^2k^2,\\
c_2 = \sqrt{B^2(2\lambda^2 + 2c\lambda-D\alpha)k^2 +
D^2\nu_0^2k^4},
\end{eqnarray}
and $C$ is normalization constant.

Without loss of generality one can put, in a numerical study,
$k=1$ and the additive noise intensity parameter $\alpha=1$ by
appropriate rescaling of the multiplicative noise intensity $D>0$,
the turbulent viscosity parameter $B>0$, the kinematic viscosity
$\nu_0>0$, and the cross correlation parameter
$\lambda$~\cite{Aringazin5}, and make a fit of $P(a)$ to the
experimental data.

The particular cases $B=0$ and $\nu_0=0$ with $\lambda$ put to
zero were studied in Ref.~\cite{Aringazin5}. As one can see from
(\ref{PLaval}), the real parameter $\lambda$ is responsible for an
asymmetry of the distribution with respect to $a\to-a$. For
$\lambda=0$ the distribution is symmetrical. Note that
$D\alpha=\lambda^2$ makes a special case, and $P(a)$ was found to
be well defined for $|a|/\langle a^2\rangle^{1/2} \leq 60$ in the
practically interesting case of negative $\lambda$ with $|\lambda|
\ll \alpha$ and $|\lambda| \ll D$. One observes also a rather
nontrivial dependence of $P(a)$ on $a$, $\lambda$ and other
parameters through $F(c)$ given by Eq.~(\ref{A4}).

\section{The conditional mean acceleration}
\label{Sec:ConditionalMean}

In this Section, we use the exact distribution (\ref{PLaval}) as a
starting point.

Since the experimental unconditional distribution $P(a)$ and the
experimental conditional distribution $P(a|u)$ at $u=0$ are
approximately of the same stretched exponential
form~\cite{Mordant0303003} we use the result of our
fit~\cite{Aringazin6} of the probability density function
(\ref{PLaval}) to the unconditional distribution
$P(a)$~\cite{Bodenschatz2}. This implied the following set of
values of the real parameters:
\begin{eqnarray}
k=1,\ \alpha=1,\ D=1.100,\ B=0.155, \nonumber \\ \nu_0=2.910,\
\lambda=-0.005,\ C=3.230.
\label{set}
\end{eqnarray}

It should be stressed that a fit to the experimental {\em
conditional} distribution $P(a|u)$ at $u=0$ would yield a
different particular set of values of the parameters. The above
fit is however justified as a first step, as we are mainly
interested in a qualitative analysis. Also, the reason that we do
not use a fit to the experimental $P(a|u)$ at $u=0$ presented in
Ref.~\cite{Mordant0303003} is an illustrative character of the
curve and that the shown range, $|a|/\langle a^2\rangle^{1/2}<15$,
is too small to capture information encoded in the long tails that
is essential for an accurate determination of the fit parameters.

Following the RIN approach, we will assume that the parameters
$\alpha$ and $\lambda$ entering (\ref{PLaval}) are stochastic and
depend on velocity fluctuations $u$. In the present paper, the
remaining parameters, $k$, $D$, $B$, and $\nu_0$, are taken to be
fixed at the fitted values given in (\ref{set}).

The exponential form of $\alpha(u)$ has been studied in
Ref.~\cite{Aringazin6} and was found to be relevant from both the
theoretical and experimental points of view. Namely, for Gaussian
distributed $u$ the positive parameter $\alpha$ is log-normally
distributed variable that corresponds to Kolmogorov 1962 (K62)
refined theory which assumes log-normal distribution of the
stochastic energy dissipation rate per unit mass, $\varepsilon$,
to which we relate $\alpha$ due to the K62 universality
hypothesis, in the inertial range.

The following remark is in order. More precisely, instead of the
stochastic energy dissipation rate it seems reasonable to use here
the stochastic energy flux through the surface of the domain of a
given spatial scale. The fluctuating energy flux is related to the
nonlinear term in the Navier-Stokes equation which dominates in
the inertial range for high-Reynolds-number flow, while the
fluctuating energy dissipation rate is associated specifically to
the dissipative scale. Under the stationarity condition (the mean
energy dissipation ${\bar\varepsilon}$ is equal to the injected
energy), the time variation of the energy contained in the given
scale is defined by the fluctuating energy flux minus fluctuating
part of the energy dissipation rate at the given scale.

From the experimental point of view, the exponential form of
$\alpha(u)$ leading to the log-normal RIN model yields the
acceleration probability density function with one free parameter
which is in a good agreement with tails of the experimental
acceleration distribution~\cite{Beck4,Aringazin5}.

Clearly, this approach builds first approximation since the
stochastic energy dissipation rate is known to follow log-normal
distribution only approximately. Also, we note that the stochastic
energy dissipation rate is positive while velocity fluctuations
$u$ are not. Hence one is motivated to seek for specific model
forms of the dependence between two stochastic variables
(monotonic Borel function), among which we choose an exponential
form for the function $\alpha(u)$ that means the same form for the
function $\varepsilon(u)$ as we will see below.

Statistical properties of the stochastic energy dissipation rate
and of the velocity fluctuations are not the same. However, they
are related to each other, particularly due to the wellknown K62
similarity hypothesis. We assume the relationship $\ln\alpha \sim
\ln\varepsilon$ in a statistical sense which is evidently
insensitive to the details related to a power law functional
dependence of $\alpha$ on $\varepsilon$, i.e., insensitive to this
particular type of nonlinearity~\cite{Beck4,Aringazin4}. This can
be viewed a manifestation of the universality. The choice of an
exponential dependence for $\alpha(u)$ and a power law dependence
for $\alpha(\varepsilon)$ implies the relationship $\varepsilon
\sim e^u$, by which we model the nonlinear relationship between
$\varepsilon$ and $u$. It should be stresses that only absolute
value of $u$ contributes the acceleration probability density
function, for normally distributed $u$. The framework for dealing
with more general situation has been recently proposed in
Ref.~\cite{Aringazin4}.

From the phenomenological point of view, the exponential form,
$\alpha(u)\sim e^{|u|}$, was found to provide appreciable increase
of the conditional acceleration variance $\langle a^2|u\rangle$
with increasing $|u|$ that meets the experimental
data~\cite{Aringazin6}.


Guided by the above observations the simplest choice is to try an
exponential dependence for $\lambda(u)$, in a phenomenological
way. Particularly, in the present paper we take
\be\label{exponenta}
\alpha(u)=e^{|u|}, \quad \lambda(u)=-0.005e^{3|u|},
\ee
which recover the values given in Eq.~(\ref{set}) at $|u|=0$. The
procedure is to refine the guess on $\lambda(u)$ depending on the
result.

Of course, the determination of functional forms for both
$\alpha(u)$ and $\lambda(u)$ is essentially (K62) phenomenological
but in general this approach is justified from the turbulence
dynamics and allows us to deal with the conditional statistics of
the fluid particle acceleration, which exhibits a quite nontrivial
behavior. It should be emphasized that the very model is justified
by the Navier-Stokes equation based LDN model~\cite{Laval}, and
the very dependence of the additive noise intensity $\alpha$ (and
the cross correlation $\lambda$) on $u$ is specific to the LDN
model, in which the additive noise was found to depend on
small-scale velocities coupled to large-scale velocities.

\begin{figure}[tbp!]
\begin{center}
\includegraphics[width=0.45\textwidth]{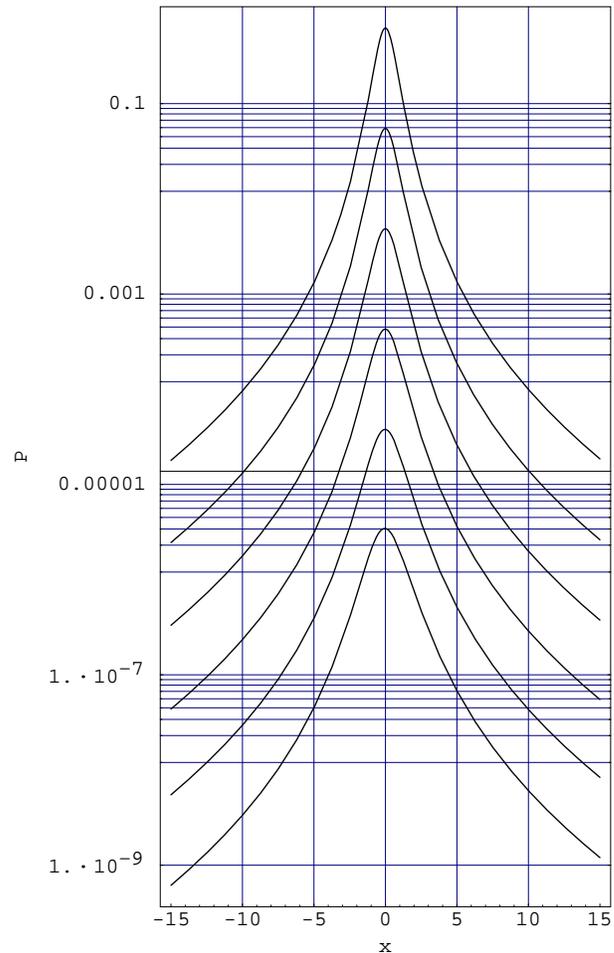}
\end{center}
\vspace{-20mm}
\caption{ \label{Fig1} The conditional acceleration probability
density function $P(a|\alpha(u), \lambda(u))$ given by
(\ref{PLaval}) for $k=1$, $\alpha=e^{|u|}$, $D=1.100$, $B=0.155$,
$\nu_0=2.910$, $\lambda=-0.005e^{3|u|}$; $u=0, 0.25, 0.50, 0.75,
1.00, 1.19$ in the rms units. The top curve is conditional on
$u=0$ while the bottom curve is conditional on $u=1.19$.
$x=a/\langle a^2 \rangle^{1/2}$ denotes normalized acceleration.}
\end{figure}

\begin{figure}[tbp!]
\begin{center}
\includegraphics[width=0.45\textwidth]{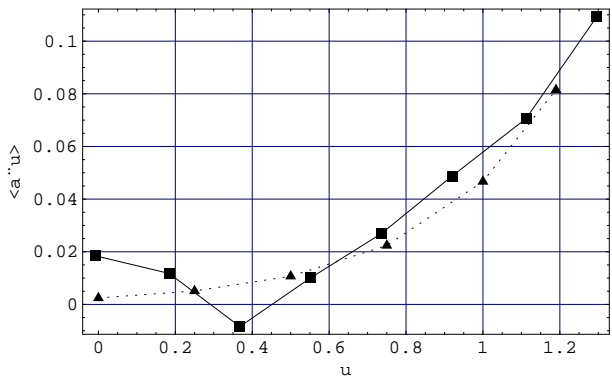}
\end{center}
\caption{ \label{Fig2} The conditional mean acceleration as a
function of standardized velocity fluctuations $u$. Triangles:
$\langle a|u\rangle/\langle a^2|0\rangle^{1/2}$ for
$\alpha=e^{|u|}$ and $\lambda=-0.005e^{3|u|}$; $k=1$, $D=1.100$,
$B=0.155$, $\nu_0=2.910$. Boxes: the experimental data on $\langle
a|u\rangle/\langle a^2\rangle^{1/2}$~\cite{Mordant0303003}. $u$ is
given in rms units.}
\end{figure}

The normalization constant $C$ in (\ref{PLaval}) will be
calculated for each value of $u$, while $C=3.230$ corresponds to
the case $u=0$. The conditional probability density function is
thus given by (\ref{PLaval}) treated in the form
$P(a|\alpha(u),\lambda(u))$ with the parameters defined by
(\ref{set}) and (\ref{exponenta}).

We are now in a position to investigate numerically how the
variation of $u$ affects the conditional statistical properties of
the component of acceleration of fluid particle.

For the normalized velocity fluctuations values $u/\langle
u^2\rangle^{1/2}=0,\, 0.25,\, 0.50,\, 0.75,\, 1.00,\, 1.19$ the
distributions $P(a|\alpha(u),\lambda(u))$ and mean accelerations
$\langle a|u\rangle/\langle a^2|0\rangle^{1/2}$ are shown in
Figs.~\ref{Fig1} and \ref{Fig2}. For convenience the curves in
Fig.~\ref{Fig1} were shifted by using repeated factor 0.1.

One observes a good qualitative correspondence of the obtained
conditional distributions plotted in Fig.~\ref{Fig1} with the
experimental curves (Fig.~6a in~\cite{Mordant0303003}). Note that
both the variance and skewness of the distribution
$P(a|u)=P(a|\alpha(u),\lambda(u))$ increase for bigger velocity
fluctuations $|u|$. While the increase of the variance [related to
the increase of $\alpha(u)$] is readily seen, the increase of the
skewness [related to the increase of $\lambda(u)$] results in
rather small change of the shape of distribution despite the fact
that $|\lambda|$ increases by about two orders of magnitude, from
$|\lambda|$=0.005 to 0.18. This change could be however readily
seen in a plot of the contribution to fourth order conditional
moment, $a^4P(a|u)$, as it is, e.g., illustrated in Fig.~4 of
Ref.~\cite{Aringazin6}. Notice that the tails of the predicted
conditional acceleration distributions in Fig.~\ref{Fig1} remain
almost the same while the central peaks become weaker for bigger
amplitudes of velocity $u$.

The obtained conditional mean acceleration $\langle
a|u\rangle/\langle a^2|0\rangle^{1/2}$ plotted in Fig.~\ref{Fig2}
is also in a good qualitative agreement with the experimental
dependence $\langle a|u\rangle/\langle a^2\rangle^{1/2}$ (Fig.~6b
in~\cite{Mordant0303003}). The mean acceleration is evidently zero
for symmetrical distribution ($\lambda=0$). One observes an
appreciable increase of the mean acceleration for bigger velocity
fluctuations $|u|$. We note that the experimental dependence of
$\langle a|u\rangle/\langle a^2\rangle^{1/2}$ on $u$ exhibits some
asymmetry with respect to $u\to -u$. This feature is not captured
by the model since we have chosen $\lambda \simeq e^{|u|}$ which
is a symmetric function.

It should be stressed that the obtained fit for the mean
acceleration is of purely {\em illustrative} character. It
demonstrates effects produced by nonzero parameter $\lambda$ which
depends on velocity fluctuations $u$, on a qualitative level. Our
primary purpose was to quantify to which extent variation of
$\lambda$ affects the conditional variance of acceleration. This
effect has been found ignorably small. Statistically homogeneous
and isotropic turbulence is characterized by zero mean
acceleration for any component in the laboratory frame of
reference. The observed nonzero conditional mean acceleration was
claimed to reflect anisotropy of the studied von Karman
flow~\cite{Mordant0303003}. It is however interesting to note that
DNS also reveals slight departures from zero. Probably this is due
to approximate character of the isotropy used in the DNS.

The reason that we make this illustration in the present paper is
that for the components of acceleration which are aligned to
trajectory of fluid particle, some skewness of the acceleration
distributions may be present. Namely, for the component pointed
transverse to the velocity, $\lambda$ is zero by construction
while for the component pointed along the velocity it is nonzero.
We remind that the acceleration is calculated due to Lagrangian
longitudinal velocity increments, $u(t+\tau)-u(t)=\tau a(t)$, in
the dissipative time-scale $\tau$. It is known that Lagrangian
longitudinal velocity structure functions of odd order are small
but not zero for inertial time-scales so that small skewness of
the velocity increments PDF should be observed. We expect that
this skewness persists in the dissipative range of time-scales
that implies (small) skewness of the acceleration distribution for
the corresponding component.

Below we make some remarks regarding numerics.

(i) We have restricted the present numerical study by the upper
value $|u|/\langle u^2\rangle^{1/2}=1.19$ because the distribution
$P(a|\alpha(u),\lambda(u))$ given by (\ref{PLaval}) turns out to
be ill-defined (discontinuous drop appears at some positive value
of $a$) for bigger normalized velocity fluctuation values. If this
is not due to a failure of the used numerical procedure at big
$|\lambda|$ in intermediate calculations, it may mean that some
adjustment of the parameters $k$, $D$, $B$, or $\nu_0$ is required
to get well-defined $P(a|\alpha(u),\lambda(u))$. Note that it is
not presumably well-defined in the entire domain of allowed
parameters values due to the presence of logarithm in (\ref{A4})
and imaginary terms. Particularly, we have found that certain
increase of the value of $B$ or $\nu_0$ implies well-defined
conditional distributions for all $|u|/\langle u^2\rangle^{1/2}$
up to 1.5 but again these become not continuous for bigger values.
Formally, the appearance of discontinuous drop of $P(a|u)$ at some
positive $a$ for big $|u|$ leads to a steeper increase
(saturation) of the conditional mean acceleration $\langle
a|u\rangle$ for big $|u|$ which is however not observed up to
$|u|/\langle u^2\rangle^{1/2}=2.5$, and the experimental
conditional distributions do not exhibit such a behavior up to
$|u|/\langle u^2\rangle^{1/2}=3$. In the physical context, the
condition that one should avoid the emerging ill-defined character
can be understood as that in addition to $\alpha(u)$ and
$\lambda(u)$ some of the other parameters should depend on $u$ in
certain way to provide well-defined distributions for any $u$.
This is important in various aspects, e.g., to provide the
integration over $u\in [-\infty,+\infty]$ to get the marginal
distribution $P(a)$. This issue is of much importance to the
present formalism but it is beyond the scope of the present paper
and can be considered elsewhere. Another possible reason of the
ill-defined character is that the conditional mean acceleration
should be very small or zero that requires much steeper increase
of the parameter $\lambda$ for bigger $|u|$.

(ii) In Fig.~\ref{Fig2} we have used $\langle a^2|0\rangle^{1/2}$
instead of $\langle a^2\rangle^{1/2}$ for normalization of the
conditional mean acceleration $\langle a|u\rangle$. This makes a
change in the overall constant factor which is obviously not of
much importance in the qualitative analysis made in the present
paper. For example, smaller value of $\langle a^2|0\rangle^{1/2}$
would shift the whole curve (triangles) in Fig.~\ref{Fig2} up in
the vertical direction. A direct numerical calculation of the mean
square, $\langle a^2\rangle= \int_{-\infty}^{\infty} da\, a^2\!
\int_{-\infty}^{\infty} du\, P(a|\alpha(u),\lambda(u))g(u)$, where
$g(u)$ is Gaussian distribution, with the above set up reveals
slow divergency which is associated to the ill-defined character
of $P(a|\alpha(u),\lambda(u))$ at big $|u|$ mentioned above.

\section{Summary}\label{Sec:Conclusions}

(i) Using the RIN approach to the LDN type one-dimensional
Langevin model of fluid particle acceleration in developed
turbulent flow, we have shown that when the cross correlation
parameter is taken in the exponential form, $\lambda \simeq
e^{|u|}$, is in a good qualitative agreement with the observed
behavior of the experimental mean acceleration conditional on
velocity fluctuations $u$, as shown in Fig.~\ref{Fig2}. This fit
is of purely illustrative character performed with the only
purpose to investigate effects produced by nonzero cross
correlation parameter $\lambda(u)$. We stress that the observed
conditional mean acceleration is related to the flow anisotropy
rather than to the correlation between stretching and vorticity
controlled by $\lambda$. Our primary purpose was to quantify to
which extent variation of $\lambda$ affects the conditional
variance of acceleration $\langle a^2|u\rangle$. This effect has
been found ignorably small as the result of the increase of
additive noise intensity $\alpha(u)$ despite the fact that
$|\lambda(u)|$ increases by about two orders of magnitude. We
encountered ill-defined character of the distribution for big
values of $|u|$ that indicates either failure of the used
numerical procedure or inappropriateness of this illustrative fit.
For homogeneous isotropic case the conditional mean acceleration
should be zero in contrast to the experimental data shown in
Fig.~\ref{Fig2}. For $\lambda=0$, the conditional acceleration
distribution is well defined for any value of $u$.

(ii) The additive noise intensity was taken to be $\alpha \simeq
e^{|u|}$ that together with $\lambda \simeq e^{|u|}$ have implied
the variation of the shape of the conditional probability
distribution function $P(a|\alpha(u),\lambda(u))$ which
qualitatively agrees with the shapes of the experimental $P(a|u)$
of the transverse component of acceleration at various values of
$u$, as shown in Fig.~\ref{Fig1}. The increase of $\alpha$ tends
to symmetrize the acceleration distribution.

(iii) Variation of $|u|$ beyond certain value was found to imply
ill-defined conditional distribution
$P(a|u)=P(a|\alpha(u),\lambda(u))$ probably because $|\lambda(u)|$
becomes comparable to the additive and multiplicative noise
intensities which case requires a more detailed study of the
nontrivial dependency of the distribution (\ref{PLaval}) on the
parameters. If this is not due to a failure of the used numerical
procedure at big $|\lambda|$ in intermediate calculations, this
problem can be cured by the assumption that some of the other
parameters of the model depend on $u$ in certain way as well, to
keep $P(a|u)$ well-defined for any $u$.

\acknowledgments{The author is grateful to M.I. Mazhitov for
stimulating discussions.
}

\end{document}